\documentclass[aapm,mph,preprint,graphicx,nofootinbib]{revtex4-1}
\usepackage[pdftex]{graphicx}
\usepackage[colorlinks=true,linkcolor=blue,citecolor=blue]{hyperref}

\usepackage[ruled, linesnumbered,noline]{algorithm2e}
\usepackage{algpseudocode} 
\usepackage{booktabs}

\makeatletter% @ becomes a letter
\def\NAT@def@citea{\def\@citea{\NAT@separator}}% Suppress spaces between citations using natbib.sty
\renewcommand{\@algocf@capt@plain}{above}% formerly {bottom}
\makeatother% @ becomes a symbol again

\DeclareGraphicsExtensions{.pdf,.jpeg,.png}
\usepackage{amsmath,amssymb,tikz}

\newcommand{\ind}[1]{\mathbf{1_{#1}}}
\newcommand{\norm}[1]{\Vert #1 \Vert}

\newcommand{\eref}[1]{(\ref{#1})}
\newcommand{\fref}[1]{Fig.~\ref{#1}}
\newcommand{\citer}[1]{Ref.~\cite{#1}}
\newcommand{\aref}[1]{Algorithm~\ref{#1}}
\newcommand{\sref}[1]{Section~\ref{#1}}
\newcommand{\tref}[1]{Table~\ref{#1}}

\usepackage{url}

\begin{document}

\def\papertitle{Superiorized method for metal artifact reduction}
\title{\papertitle}

\author{T.~Humphries}
\email[Email: ]{thumphri@uw.edu}
\affiliation{School of STEM, University of Washington Bothell}
\author{J.~Wang}
\affiliation{School of STEM, University of Washington Bothell}

%\author{A.~Faridani}
%\affiliation{Department of Mathematics, Oregon State University}

\date{\today}

\begin{abstract}	% <= 300 words
\textbf{Purpose:}  Metal artifact reduction (MAR) is a challenging problem in computed tomography (CT) imaging. A popular class of MAR methods replace sinogram measurements that are corrupted by metal with artificial data, typically generated from some combination of interpolation along with other heuristics. While these ``projection completion'' approaches are successful in eliminating severe artifacts, secondary artifacts may be introduced by the artificial data. In this paper, we propose an approach which uses projection completion to generate a prior image, which is then incorporated into an iterative reconstruction algorithm based on the superiorization framework. The rationale is that the image produced by the iterative algorithm can inherit the desirable properties of the prior image, while also reducing secondary artifacts.

\textbf{Methods:} The prior image is reconstructed using normalized metal artifact reduction (NMAR), a popular projection completion approach. The iterative algorithm is a modified version of the simultaneous algebraic reconstruction technique (SART), which reduces artifacts by incorporating a polyenergetic forward model, least-squares weighting, and superiorization. The penalty function used for superiorization is a weighted average between a total variation (TV) term and a term promoting similarity with the prior image, similar to penalty functions used in prior image constrained compressive sensing (PICCS). Because the prior is largely free of severe metal artifacts, these artifacts are discouraged from arising during iterative reconstruction; additionally, because the iterative approach uses the original projection data, it is able to recover information that is lost during the NMAR process.

\textbf{Results:} We perform numerical experiments modeling a simple geometric object, as well as several more realistic scenarios such as metal pins, bilateral hip implants, and dental fillings placed within an anatomical phantom. The proposed iterative algorithm is largely successful at eliminating severe metal artifacts as well as secondary artifacts introduced by the NMAR process, especially lost edges of bone structures in the neighborhood of the metal regions. In one case modeling severe photon starvation, the NMAR algorithm is found to provide better results.

\textbf{Conclusion:} The proposed algorithm is effective in applying the superiorization methodology to the problem of MAR, providing better results than both NMAR and a purely total variation-based superiorization approach in nearly all cases.

\end{abstract}

\maketitle

%\linenumbers\modulolinenumbers[5]

\section{Introduction}

Artifacts caused by metal objects, such as prostheses, rods, screws, and fillings, are a well-known source of image artifacts in computed tomography (CT). These artifacts are caused by numerous physical factors, including beam hardening, increased measurement noise, photon starvation, partial volume and exponential edge gradient effects, and scatter~\cite{DNDMS98}. They typically appear as dark streaks or bands between metal objects, as well as thin, alternating dark and light streaks emanating from these objects. These artifacts tend to be dramatic and may obscure important features of the image.

Despite the long history of proposed metal artifact reduction (MAR) techniques (see~\citer{gjest16} for a recent review), it remains a challenging problem. Dual-energy systems are able to significantly reduce metal artifacts~\cite{BDNRBJ11}, but require special equipment. On conventional systems, a common approach, often referred to as projection completion, regards measurements that have passed through a metal object as unreliable, and replaces them with artificially generated data. Typically, a preliminary, artifact corrupted image is first reconstructed from the raw sinogram to identify metal regions in the image space, followed by a simulated forward projection to identify the so-called metal trace in the sinogram domain. Once data inside the metal trace have been replaced, an image is reconstructed from the modified data using a standard technique such as filtered backprojection (FBP). 

Early projection completion methods~\cite{GP81,KHE87} used linear interpolation (LI) to replace the metal trace within each column of the sinogram. While the LI approach is effective in removing the most severe metal artifacts, it also produces secondary artifacts, due mainly to the loss of edge information about other structures (e.g. bone or air pockets) where their traces intersect with the metal trace~\cite{MB09}. A popular way to address this issue is to use the preliminary image to not only identify the metal trace, but also generate a prior image, which omits metal but retains information about the other structures such as bone. A number of authors~\cite{BS06,PKBK09,KCWM12,LFN09,VS12,ZYJYJ13,THZ13,ZY18} propose replacing the metal trace with the corresponding measurements obtained from simulated forward projection of the prior, rather than LI data. Alternatively, in normalized metal artifact reduction (NMAR) methods, the prior is used to normalize the projections~\cite{MB09,MRLSK10,MRLSK12} before a sinogram correction using LI; the sinogram is subsequently ``de-normalized'' to reintroduce lost edge information. The prior image is typically generated using tissue classification of an image reconstructed using FBP~\cite{BS06,PKBK09,MRLSK10,KCWM12} or by iterative reconstruction with regularization to suppress artifacts~\cite{LFN09,VS12,ZYJYJ13,THZ13}; one recent work~\cite{ZY18} has also proposed generating the prior using a convolutional neural network.

Some approaches based on projection completion employ additional heuristics to reduce secondary artifacts. \citer{WCSLX04} proposes segmenting bone from the preliminary reconstruction (replacing bone pixels with smoothed soft tissue values) and then reintroducing bone to the image after an interpolation-based projection completion is performed. \citer{WK04} uses a distance-dependent spatial weighting to combine the LI image with one processed using multi-dimensional adaptive filtering to reduce noise. \citer{BF11} begins with an LI reconstruction and then performs several iterations in which the reconstruction is forward projected, averaged with experimental data using spatial weighting, then reconstructed again. Other methods retain edge information by performing projection completion in the Fourier~\cite{KWB12} or wavelet~\cite{ZRWWB00,MARZ13,BC17} domains, rather than on the original sinogram. Another recent approach~\cite{GSYXC18} uses a convolutional neural network to combine information from an NMAR image and the uncorrected image. Finally, projection completion can also be performed directly in the sinogram domain without reconstruction of a preliminary image~\cite{VJVG10}, avoiding issues of mismatch between the simulated forward projection and true system geometry.

An alternative to projection completion is to use the original data and reconstruct the image using an iterative algorithm. Iterative algorithms, while more computationally expensive than FBP, have the advantage of being able to accurately model physical effects contributing to metal artifacts, such as beam hardening and noise. They also may be able to solve the exterior problem (where the metal-corrupted data are ignored rather than replaced) with the inclusion of regularization terms to mitigate the ill-posedness of the problem. \citer{WSOV96} applied both the algebraic reconstruction technique (ART) and expectation maximization (EM), finding that both outperformed FBP reconstruction. Later work has incorporated polyenergetic modeling into the EM framework~\cite{DNDMS01,VN12}, applied EM to projection completed data~\cite{OB07}, used ART in conjunction with total variation (TV) penalties~\cite{RBFK11,ZWX11} and used optimization-based reconstruction with machine-learned regularization~\cite{BC17}.  In~\citer{CYSTS19}, the authors simultanously estimate the image and the mismatch between polyenergetic and (idealized) monoenergetic data, using a regularized least squares approach incorporating a prior image and several different regularizers.

Projection completion approaches and iterative methods each have advantages and disadvantages. Projection completion is especially effective in removing severe streaking artifacts from images, but the use of artificially generated data carries an inherent risk of creating new artifacts. Iterative methods can make use of the original projection data, but the poor quality of the metal contaminated measurements make obtaining a high quality image difficult. In this paper we combine projection completion with an iterative method to attempt to mitigate these issues. We use the NMAR method to construct a prior image which is free of severe artifacts, but may contain some secondary artifacts. This image is then used both as the initial estimate and as a prior in an iterative method which reconstructs an image from the original projection data. The iterative method incorporates the superiorization methodology~\cite{HGDC12} with a secondary objective guided by the prior image as well as a total variation (TV) minimization term. Numerical experiments indicate that the proposed algorithm is successful in eliminating both severe artifacts due to streaking, as well as secondary artifacts introduced during the NMAR process.

\section{Methodology}

\subsection{Mathematical model}

We consider a two-dimensional attenuation distribution, $\mu(y, E): \mathbb{R}^2 \times \mathbb{R} \to \mathbb{R}$, which depends on position, $y$, and the energy of the incident X-ray beam, $E$. If the X-ray beam is monoenergetic with energy $E_0$, the idealized measurement along a line $j$ is modeled as
\begin{equation}
\widehat{I}_j = I_0 \exp \left( -\int_j \mu(y, E_0) \: d y \right), \label{E:mono1}
\end{equation}
where $I_0$ is the initial intensity of the beam and $\widehat{I}_j$ is the (idealized) intensity measured by the detector, in counts per second. The line integral, $\int_j \mu(y, E_0) \: dy$, which is a sample from the Radon transform of $\mu(y, E_0)$, can be obtained by log-transforming the data,
\begin{equation}
\ln \left(I_0 / \widehat{I}_j\right) = \int_j \mu(y, E_0) \: d y.  \label{E:mono2}
\end{equation}
Discretizing $\mu$ as an image with $K$ pixels, and taking measurements along $J$ lines, then yields a system of linear equations

\begin{equation}
A x = b, \label{E:mono3}
\end{equation}
where $x \in \mathbb{R}^K$ is the discretized image of $\mu$, $b \in \mathbb{R}^J$ is the line integral data, and $A \in \mathbb{R}^{J \times K}$ is the system matrix whose $(j,k)$th element is the length of intersection of the $j$th line with the $k$th pixel of $x$. 

Clinical CT systems, however, generate polyenergetic X-ray beams, with a spectrum of energies typically ranging from zero up to roughly 150 keV. In this instance, the measurement model is expressed as
\begin{equation}
\widehat{I}_j = \int S(E)   \exp \left( -\int_j \mu(y, E) \: d y \right) d E, \label{E:poly1}
\end{equation}
where $S(E)$ is the beam spectrum. Log-transforming the data and applying the mean value theorem for integrals then gives
\begin{align}
\ln \left(I_0 / \widehat{I}_j\right) &=    \int_j \mu(y, E_j) \: d y, \label{E:poly2}
\end{align}
where $I_0 = \int S(E) \: dE$, and $E_j$ is an unknown energy in the range of the spectrum, depending on the path $j$. The inconsistency with the monoenergetic model~\eref{E:mono2} leads to so-called beam hardening artifacts~\cite{BD76}, including cupping and streaking. From the physical perspective, as the polyenergetic beam passes through the object, photons with lower energy are attenuated at a higher rate than photons with greater energy, causing the spectrum of the beam to ``harden'' as it becomes skewed towards higher energies. Metal objects induce particularly severe beam hardening artifacts due to their high attenuation coefficients, particularly at the low end of the energy spectrum.

Another contributing factor to metal artifacts is increased image noise. Due to the stochastic nature of X-ray interactions with matter, the measurement along line $j$ is typically modeled as a Poisson random variable $I_j$, with mean equal to $\widehat{I}_j$. The measurement therefore has a signal-to-noise ratio (SNR) of $\sqrt{\widehat{I}_j}$; i.e., the SNR deterioriates as $\widehat{I}_j$ becomes small. This situation occurs if the initial beam intensity $\int S(E) \: dE$ is small, or if the line integral through $\mu(y,E)$ is large. Measurements passing through metal objects thus tend to be very noisy, degrading image quality. In the extreme case $\widehat{I}_j$ may be so small that $I_j = 0$ (no counts are detected along the line), resulting in an effective attenuation measurement of infinity. 

\subsection{Beam hardening and metal artifact reduction}

A standard approach to reduce beam hardening artifacts is to perform water correction (sometimes known as soft tissue correction) on the measured data~\cite{JS78}. Water correction (\aref{A:softtissue}) simulates a monoenergetic measurement $m_j$ from the corresponding polyenergetic measurement $I_j$ via a two step process. First, the effective length, $T_j$, of water through which the polyenergetic beam would have to pass to generate $I_j$ is estimated. This can be accomplished by solving a nonlinear equation (Line 3 of \aref{A:softtissue}, with $\mu_w(E)$ denoting the attenuation coefficient of water) or, more commonly, by interpolating from a table of measured values based on known thicknesses of water. Once $T_j$ is known, the corresponding measurement $m_j$ at some reference energy $E_0$ can be obtained straightforwardly.

\begin{algorithm}
\caption{Water correction.}\label{A:softtissue}  

\Indpp

{\em Given:} Polyenergetic measurements $I_j$, $j \in [1,J]$.

\medskip

\For {$j = 1$ to $J$}{
~~~Solve $I_j = \int S(E) \exp ( -\mu_w(E) T_j) \: dE$ for $T_j$\;
~~~Set $m_j = I_0 \exp (-\mu_w(E_0) T_j)$\;
}
\Return $m_j, j \in [1,J]$.

\end{algorithm}

Water correction essentially assumes that the object consists only of water or water-like materials. This assumption breaks down if the object contains dense material like bone, contrast agents, or metal, whose attenuation curves do not behave like scaled versions of the attenuation curve of water. Water correction is therefore effective in reducing cupping artifacts, but not the so-called second-order artifacts caused by those materials. Correcting for these artifacts typically requires more advanced data corrections~\cite{JS78, JR97, KMPK10} or iterative methods which directly model polyenergetic X-rays~\cite{DNDMS01,EF02, VVDBS11,LS14}. %The iterative reconstruction method we propose in this paper is based on one of the latter approaches~\cite{LS14}, which we describe in the next section.

As discussed in the Introduction, a standard method of metal artifact reduction (MAR) is to replace metal-contaminated measurements with linearly interpolated data~\cite{GP81,KHE87}. Pseudocode for this method, which we denote as MAR, is provided in \aref{A:MAR}. A preliminary image $x_{uncorr}$ is reconstructed from the water corrected sinogram, and the image is segmented into non-metal and metal parts using thresholding. The metal index $\ind{m}$ is forward projected to obtain the metal trace in the sinogram domain. The sinogram is then processed columnwise (i.e. with respect to each projection angle) to replace measurements corresponding to the metal trace with their interpolated values. An image can then be reconstructed from the corrected sinogram, and the metal objects can be re-inserted if desired.

\begin{algorithm}
\caption{MAR method}\label{A:MAR}

\Indpp

{\em Given:} Water-corrected sinogram, $b_{mono}$, and system matrix, $A$.

\medskip 

Reconstruct image $x_{uncorr}$ from $b_{mono}$\;
Segment $x_{uncorr}$ to obtain metal index $\ind{m}$\;
Set $trace = A\:\ind{m}$ (forward projection)\;
Set $b_{mono}=0$ wherever $trace > 0$ \;
\For{every column of $b_{mono}$}{
Fill in zero values by linear interpolation to obtain corresponding column of $b_{MAR}$\;
}
\Return $b_{MAR}$
\end{algorithm}

The MAR method reduces severe artifacts caused by metal, but produces secondary artifacts, as the interpolation results in the loss of edge information about other structures. The normalized metal artifact reduction (NMAR) method~\cite{MRLSK10} (\aref{A:NMAR}) addresses this issue through the use of a prior image. Following the preliminary reconstruction, $x_{uncorr}$ is segmented into air, soft tissue, bone, and metal. The prior image is created by setting regions containing air and soft tissue to their respective attenuation values at the reference energy. Regions containing bone are assigned the corresponding attenuation coefficients from $x_{uncorr}$, to accurately capture the varying attenuation of bone. The values assigned to the metal regions are unimportant~\cite{MRLSK10}; we use soft tissue for simplicity. The prior image is then forward projected, and the resulting sinogram $b_{prior}$ is divided elementwise ($\oslash$) into the original sinogram to normalize it. The MAR algorithm is then applied to the normalized sinogram, followed by an elementwise multiplication ($\otimes$) with $b_{prior}$ to ``de-normalize'' the sinogram. The de-normalization step effectively recovers edge information about bone and other structures that may be lost during the MAR step~\cite{MB09, MRLSK10}. 

\begin{figure}
\includegraphics[width=0.6\linewidth]{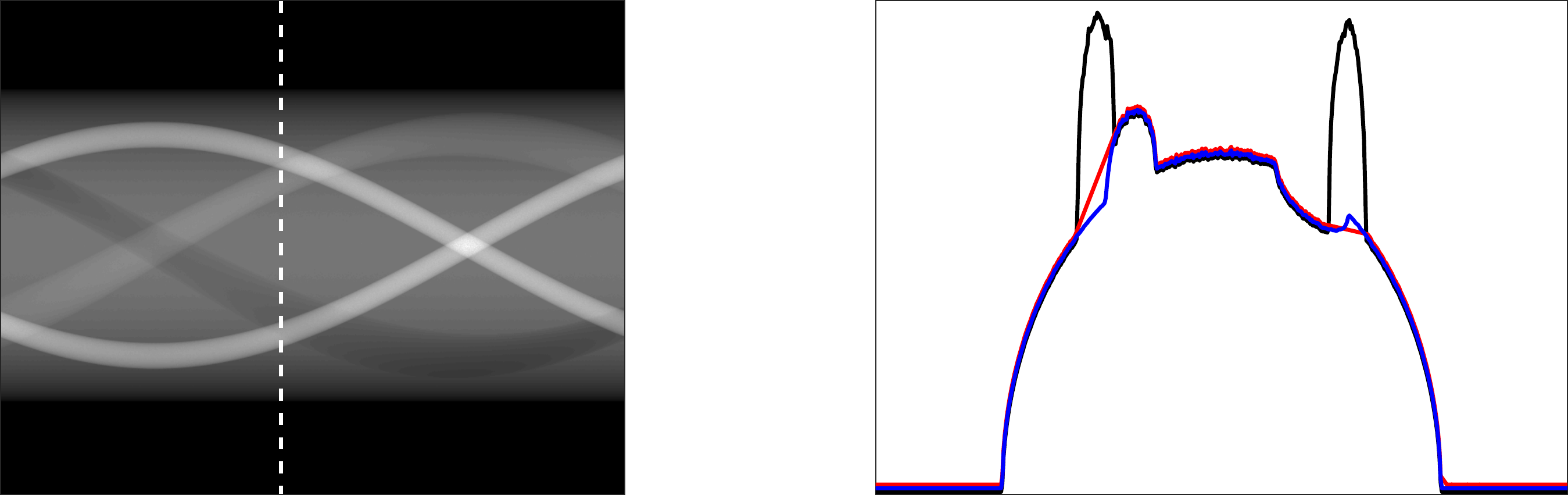}

\bigskip

\includegraphics[width=0.9\linewidth]{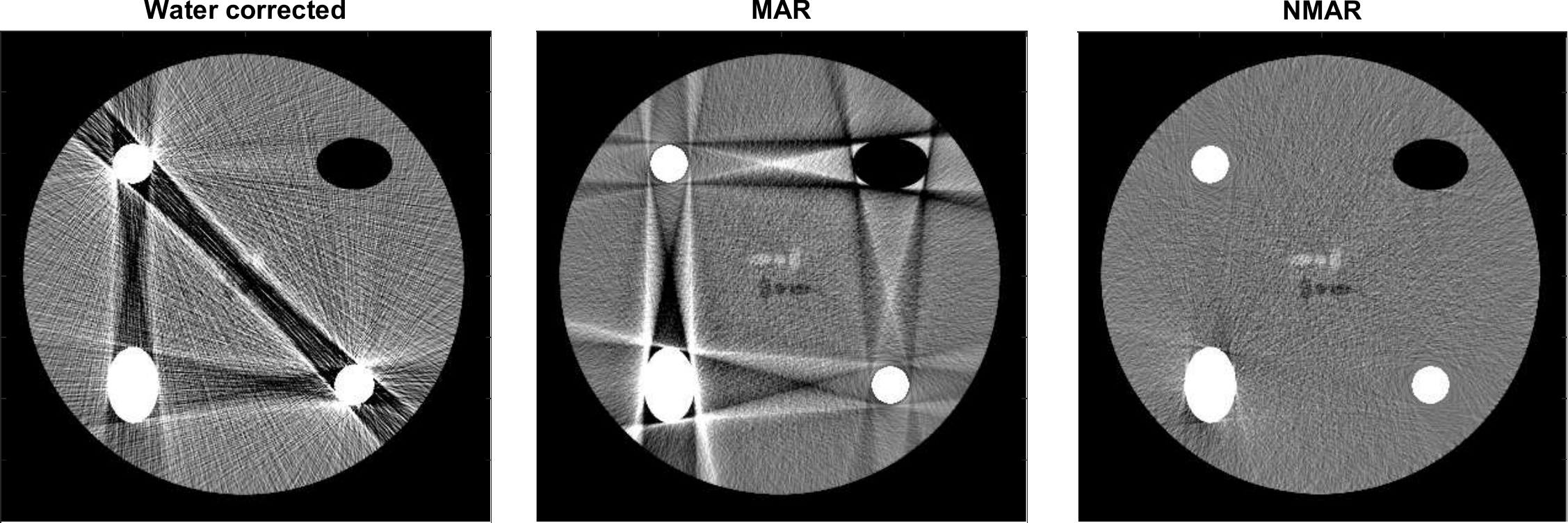}

\caption{Illustration of MAR and NMAR methods. Top row: Left: sinogram of a simple mathematical phantom including metal objects (small circular regions) as well as bone (oblong region in bottom right) and air pocket (top right). Right: Profile along white dashed line in sinogram, showing water corrected sinogram (black line), MAR correction (red line) and NMAR correction (blue line). Profiles have been vertically offset to highlight differences. Bottom: Images reconstructed from water corrected (left), MAR corrected (center) and NMAR corrected (right) sinograms using filtered backprojection. Phantom specifications are given in \sref{S:results}.}\label{F:MARvsNMAR}
\end{figure}

\fref{F:MARvsNMAR} highlights the differences between MAR and NMAR. In the sinogram profiles (top right) we can observe that the interpolated profile used by MAR (red line) does not accurately capture the traces of the bone and air pocket, while the NMAR method does. While the MAR correction (bottom center) removes the large streak between the two metal objects that appears in the water corrected (bottom left) image, secondary streaking artifacts are created between the metal objects and the bone and air pockets. NMAR is able to remove virtually all artifacts in this simple experiment, though a mild artifact persists surrounding the bone region.

\begin{algorithm}
\caption{NMAR method}\label{A:NMAR}

\Indpp

{\em Given:} Water-corrected sinogram, $b_{mono}$, system matrix, $A$, reference energy $E_0$.

\medskip

Reconstruct image $x_{uncorr}$ from $b_{mono}$\;
Segment $x_{uncorr}$ to obtain indices $\ind{a}, \ind{s}, \ind{b}$ and $\ind{m}$ (air, soft tissue, bone, metal)\;
Set $x_{prior} = \mu_a(E_0)$ wherever $\ind{a} = 1$\; 
Set $x_{prior}= \mu_s(E_0)$ wherever $\ind{s} \cup \ind{m} = 1$\; 
Set $x_{prior}=  x_{uncorr}$ wherever $\ind{b} = 1$\;
$b_{prior} = A x_{prior}$ \;
$b_{norm} = b_{mono} \oslash b_{prior}$ \;
$b_{MAR} = MAR(b_{norm}, A)$ \;
$b_{NMAR} = b_{MAR} \otimes b_{prior}$\;
\Return $b_{NMAR}$
\end{algorithm}

\subsection{Iterative reconstruction}

Standard iterative reconstruction methods are based on solving the linear system $Ax = b$ as described in \eref{E:mono3}. Our approach is based on a block-iterative variant of the simulataneous algebraic reconstruction technique (SART)~\cite{AK84}, as described in \citer{CE02}. We first define diagonal matrices $D$ and $M$ as:
\begin{align*}
D &\in \mathbb{R}^{K \times K},~ D_{kk} = 1 \biggl/ \sum_{i=1}^J |a_{ik}|\\
M &\in \mathbb{R}^{J\times J},~M_{jj} = 1 \biggl/ \sum_{i=1}^K |a_{ji}|
\end{align*}
To implement block iteration, the columns of the sinogram are evenly partitioned into $N_w$ subsets, indexed by $w$; for example, if $N_w = 5$, then $w=1$ would correspond to the first, sixth, eleventh, etc. columns, $w=2$ to the second, seventh, twelfth, and so on.This accelerates the convergence of the algorithm. One iteration of the block iterative SART (BI-SART) algorithm is then given by
\begin{align}
x^{(i+1)}    &= \mathbf{Q} \mathbf{B}_{N_w} \dots \mathbf{B}_2 \mathbf{B}_1 (x^{(i)}), \text{where}\label{E:BISART1}\\
\mathbf{B}_w (x) &= x  - D_{w} (A_{w})^T M_{w}\left[ A_{w} x - b_{w} \right],  \label{E:BISART2}\\
(\mathbf{Q} x)_k &= \max\{0, x_k\},~~k \in [1, K]. \notag
\end{align}
The subscript $w$ indicates that only rows of $A$ and $b$ corresponding to the measurements in $w$ are used, including when forming the matrices $D_w$ and $M_w$. The operator $\mathbf{Q}$ ensures non-negativity of the solution after every iteration.

The BI-SART algorithm accounts for neither the statistical uncertainty in the measured data, nor the polyenergetic nature of the X-ray measurements. Additionally, images reconstructed by BI-SART may be of poor quality when the data are noisy. In \citer{HG17}, we introduced three enhancements to the BI-SART algorithm to address these issues, which are especially important in the context of metal artifact reduction. We summarize these enhancements below:

\subsubsection{Polyenergetic forward model}

To account for the polyenergetic measurements, we adopt the polyenergetic SART (pSART) method~\cite{LS14b}. pSART defines a polyenergetic forward projection operation $\mathcal{P}$ using linear interpolation between tabulated attenuation curves for known basis materials such as soft tissue, bone, and metal. Specifically, letting $x$ denote the attenuation image at the reference energy $E_0$, we define
\begin{align}
&\mu(x_k, E) = \notag \\ 
&~ \frac{ [\mu_{m+1}(E_0) - x_k] \mu_{m}(E) +  [x_k - \mu_{m}(E_0) ] \mu_{m+1}(E)}{\mu_{m+1}(E_0) - \mu_{m}(E_0)}, \label{E:mu_interp}
\end{align}

where $\mu_m(E)$ and $\mu_{m+1}(E)$ are the linear attenuation coefficient (LAC) curves of the two basis materials whose LACs at $E_0$ bracket the pixel value $x_k$. For example, if $x_k$ has an LAC halfway between that of soft tissue and bone at $E_0$, its LAC is assumed to be halfway between that of soft tissue and bone at all other energies as well. The polyenergetic forward projection operation is then defined as
\begin{align}
[\mathcal{P} (x)]_j &= \ln \left[  \sum_{h=1}^{N_h} S_h \biggl/ \sum_{h=1}^{N_h} S_h \exp \left( - a_j \mu (x, E_h)  \right) \right],\label{E:fp_poly}
\end{align}

where $S_h$ is a discretization of the continuous beam spectrum $S(E)$ into $N_h$ energy levels, and $a_j$ is the $j$th row of $A$ (cf. \eref{E:poly1} and \eref{E:poly2}). The pSART iteration is achieved by simply replacing the monoenergetic forward projection $Ax$ in SART with $\mathcal{P}(x)$ in \eref{E:BISART2}.:

\begin{equation}
\mathbf{B}_w (x) = x  - D_{w} (A_{w})^T M_{w}\left[ \mathcal{P}_{w}( x) - b_{w} \right]. \label{E:pSART}
\end{equation}
While pSART has not been proven to converge in general, it has been shown to be effective in reducing beam hardening artifacts in numerical experiments~\cite{LS14b,H15,HWF17}.

\subsubsection{Weighted least squares} To model measurement uncertainty, a weighted least squares approach (WLS)~\cite{SC00} can be employed. We define the diagonal weighting matrix as
\begin{equation}
W \in \mathbb{R}^{J\times J},~W_{jj} = I_j, \label{E:weight_matrix}
\end{equation}

We can then apply BI-SART to the system $W^{\frac{1}{2}} Ax = W^{\frac{1}{2}} b$, which assigns proportionally higher weight to the less noisy measurements. Incorporating the polyenergetic forward projection as well, the modified iteration is given by
\begin{align}
\mathbf{B}_w (x) &= x  - D'_{w} (A_{w})^T M_{w}\left[ W_w^{\frac{1}{2}}\left( \mathcal{P}_{w}( x) - b_{w} \right) \right],\label{E:WLS}
\end{align}
where $\displaystyle D'_{kk} = 1 \biggl/ \sum_{i=1}^J \left \vert (W^{\frac{1}{2}} A)_{ik}\right\vert$ \footnote{Matrix $M$ does not need to be modified, as the factor of $W^{\frac{1}{2}}$ is cancelled during multiplication with $(W^{\frac{1}{2}} A)^T$.}.
%The matrix $M$ does not need to be modified, since multiplying $A$ by $W^{\frac{1}{2}}$  has the effect of multiplying $M$ by $W^{-\frac{1}{2}}$, which is then canceled by the multiplication with $(W^{\frac{1}{2}} A)^T$. 

\subsubsection{Superiorization}

Superiorization~\cite{HGDC12} is an optimization heuristic in which solutions generated by an iterative algorithm are perturbed in every iteration, to improve the quality of the solution in some respect. For example, if \eref{E:BISART1} is taken to be the basic iterative algorithm with $\mathbf{B}_w$ defined in \eref{E:WLS}, then the superiorized version is of the form

\begin{equation}
x^{(i+1)}    = \mathbf{Q} \mathbf{B}_{N_w} \dots \mathbf{B}_2 \mathbf{B}_1 (x^{(i)}, + \beta^{(i)} v^{(i)}),\label{E:BISARTsup}\\
\end{equation}
where $\beta^{(i)} > 0$, $\sum_i \beta^{(i)} < \infty$, and the $v^{(i)}$ are a sequence of bounded perturbation vectors. The key result underlying superiorization is that if the basic algorithm is {\em perturbation resilient}, the superiorized version is able to find a solution that is equally satisfactory with respect to satisfying the constraints of the problem ($\mathcal{P}(x) = b$ in our case), though typically this requires more iterations. The perturbation vectors are usually chosen to be nonascending directions of some penalty function $\phi(x)$ at the current iterate, for example, $v^{(i)} = - \nabla \phi(x^{(i)}) / \norm{\nabla \phi(x^{(i)})}$. Thus, the solution found by the superiorized algorithm can be expected to be superior with respect to $\phi$, while equally compatible with the constraints. The methodology has been featured in a special edition of {\em Inverse Problems}~\cite{CHJ17}, and a continously updated online bibliography is maintained at \url{http://math.haifa.ac.il/YAIR/bib-superiorization-censor.html}.

All three of the proposed enhancements (polyenergetic forward projection, weighted least squares, and superiorization) can be implemented independently of one another. In~\cite{HWF17} we found that polyenergetic forward projection and TV superiorization could be used to eliminate beam hardening artifacts as well as artifacts due to sparse-view and limited-angle data. In~\cite{HG17}, we found that including all three enhancements in the reconstruction algorithm provided the best results when metal artifacts were present. Pseudocode for this algorithm is presented in \aref{A:AlgwpSART-sup}. In the implementation of the algorithm, we allow for a total of $N$ perturbations to be applied in every iteration, to further improve the solution with respect to the penalty function. (This is of course equivalent to a single perturbation, albeit one that cannot be determined {\em a priori}). The step sizes decrease geometrically with rate $0 < \gamma < 1$ to ensure that the perturbations are summable. 

In \cite{HG17}, total variation (TV) was used as the penalty function employed by superiorization:
\begin{align}
&\phi_{TV}(x) = \notag \\
&~\sum_{m,n} \sqrt{ (x_{m+1,n}-x_{m,n})^2 + (x_{m,n+1}-x_{m,n})^2 + \epsilon^2},\label{E:TV}
\end{align}
where $x_{m,n}$ denotes the pixel in the $m$th row and $n$th column of the image $x$. The small parameter $\epsilon$ is introduced to avoid singularity of $\nabla \phi_{TV}$ at points where the image is piecewise constant. While the algorithm, which we denoted as wPSART-TV (weighted polyenergetic SART with TV superiorization) was effective in reducing metal artifacts, we found that it was difficult to remove strong streaking artifacts between metal objects; for example, in the case of a bilateral hip implant. This was especially true at low count rates, where photon starvation was a significant factor. This motivates the development of the hybrid algorithm, discussed in the next section.

\begin{algorithm}
\SetAlgoHangIndent{1cm}
\caption{Block-iterative, weighted, polyenergetic SART with superiorization}\label{A:AlgwpSART-sup}

\Indpp

{\em Given:} Log-transformed polyenergetic sinogram $b_{poly}$, system matrix $A$, number of subsets $N_w$, reference energy $E_0$, spectrum $S$, tabulated attenuation curves $\mu_m(E)$, initial image $x_0$, stopping criteria $\varepsilon$ and/or $i_{max}$, superiorization parameters $\gamma$, $N$.

\medskip

$\ell = -1$\;
$i = 0$\;
\While{$\norm{\mathcal{P}\left(x^{(i)}\right) - b_{poly}}_2 \geq \varepsilon$ {\bf and } $i < i_{max}$}{
\Indpp
$n = 0$ \;
$x^{(i,n)} = x_i$\;
    \While{$n < N$}{
    \Indpp
    $v^{(i,n)} = -\nabla \phi (x^{(i,n)})~/~ \norm{ \nabla \phi(x^{(i,n)})}_2$\;
    \While{\bf true}{
        \Indpp
        $\ell = \ell+1$\;
        $\beta^{(i,n)} = \gamma^\ell$\;
        $z =  x^{(i,n)} + \beta^{(i,n)} v^{(i,n)}$\;
        \If {$z \in \Omega$ and $\phi(z) \leq \phi(x^{(i)})$}{
           \Indpp
            $x^{(i,n+1)} = z$\;
            {\bf break}\;
           \Indmm
            }
           \Indmm
        }
        $n = n + 1$\;
        \Indmm 
    }
$x^{(i+1)} = \mathbf{Q} \mathbf{B}_{N_w} \dots \mathbf{B}_2 \mathbf{B}_1 (x^{(i,N)})$, where $\mathbf{B}_w (x) = x  - D'_{w} (A_{w})^T M_{w}\left[ W_w^{\frac{1}{2}}\left( \mathcal{P}_{w}( x) - b_{poly,w} \right) \right]$\;
$i = i + 1$\;
\Indmm
}
\Return $x^{(i)}$
\end{algorithm}

\subsection{Prior image constrained superiorized algorithm}

Our prior image constrained superiorized algorithm (\aref{A:PICS}), which we denote as wPSART-PICS, consists of several steps. First, the water corrected sinogram is corrected using NMAR to obtain a metal corrected sinogram. A modified version of \aref{A:AlgwpSART-sup} is then applied to this sinogram to reconstruct a prior image, $x_{prior}$. In the modified version of the algorithm, monoenergetic forward projection (i.e. multiplication by $A$) is used instead of $\mathcal{P}$, and the weighting matrix $W$ is omitted, since the data have been water corrected and the noisy metal trace removed by NMAR. Total varation~\eref{E:TV} is used as the penalty function for superiorization, in order to generate a smooth prior image.

The prior image is then incorporated into a second penalty function:
\begin{equation}
\phi_{PI}(x) = \alpha \phi_{TV} (x) + (1-\alpha) \phi_{TV} (x - x_{prior}), \label{E:TVP}
\end{equation}
where $\alpha \in [0,1]$ controls the weighting of the two terms. This type of penalty is used in the prior image constrained compressed sensing (PICCS) approach described in~\cite{CTL08, LTC12}; we use the term ``superiorization'' instead of ``compressed sensing'' as it more accurately describes our method. \aref{A:AlgwpSART-sup} is then applied with the original polyenergetic projection data and $\phi_{PI}(x)$ as the penalty function. The prior image is also used as the initial estimate, as it is expected to be a good approximation to the true image.

\begin{algorithm}
\caption{wPSART-PICS algorithm}  \label{A:PICS}

\Indpp

{\em Given:} Water corrected sinogram $b_{mono}$, all inputs needed for \aref{A:AlgwpSART-sup}.

\medskip

Apply NMAR to $b_{mono}$ to obtain $b_{NMAR}$\;
Use \aref{A:AlgwpSART-sup} (modified) with $b_{NMAR}$ as sinogram to reconstruct prior image $x_{prior}$\;
Set $\phi_{PI}(x) = \alpha \phi_{TV} (x) + (1-\alpha) \phi_{TV} (x - x_{prior})$\;
Use \aref{A:AlgwpSART-sup} with $b_{poly}$ as sinogram, $x_0 = x_{prior}$, and $\phi_{PI(x)}$ as penalty function to reconstruct $x_{final}$\;
\Return $x_{final}$
\end{algorithm}

By penalizing the TV of the difference between the reconstructed image and the prior, $\phi_{PI}(x)$ effectively promotes similar edge structure between the two images. Our motivation for using $\phi_{PI}(x)$ in this application is that if the NMAR-based prior is largely free of severe streaking artifacts, these artifacts will also be penalized in the image being reconstructed. Since the final step of \aref{A:PICS} uses the original polyenergetic data, however, it should be possible to recover image details that are lost due to the interpolation performed by NMAR.

\section{Results}\label{S:results}

We validated our approach using several numerical experiments. The wPSART-PICS algorithm was implemented in Matlab using the Michigan Image Reconstruction Toolbox (MIRT)~\cite{MIRT} to simulate the CT spectrum, material attenuation curves, and generate the system matrix $A$ for iterative reconstruction. We implemented our own version of NMAR in Matlab as well, using the built-in Image Processing Toolbox methods to segment the CT image.

\subsection{Simple phantom experiments}

We first generated the simple mathematical phantom shown in \fref{F:MARvsNMAR} to fine tune the algorithm parameters, before studying more realistic data. The phantom is $400 \times 400$ pixels with a pixel size of 0.75 mm. It consists of a background region of soft tissue, two small circular regions containing titanium, and a larger oblong region containing bone. The linear attenuation coefficients of soft tissue, bone and titanium at the reference energy $E_0$ of 70 keV are $0.203$, $0.494$, and 2.44 cm$^{-1}$, respectively. The phantom also consists of an air pocket and several low contrast features (modeled after the Shepp-Logan phantom) in the center of the object. The low-constrast features have attenuation values of $\pm 5 \%$ relative to the background, and are heavily obscured by metal streaking artifacts if no metal artifact correction is performed (see \fref{F:MARvsNMAR}).

We analytically computed 720 parallel-beam views over a 180$^\circ$ arc around the phantom. A 130 kVp spectrum was simulated to generate polyenergetic data. Poisson noise was subsequently added to the data, proportionally to initial count rates of $I_0 = 1.0 \times 10^5$, $2.0 \times 10^5$, $5.0 \times 10^5$, and $1.0 \times 10^6$. To reconstruct the data using \aref{A:PICS}, we ran 24 iterations of the modified algorithm (Line 3) to generate $x_{prior}$. \aref{A:AlgwpSART-sup} was then run for 32 iterations to produce a final image. In both instances, $N_w = 12$ subsets of projection data were used. The parameters controlling the gradient descent steps within the superiorized algorithm were set to $\gamma = 0.9995$, $N = 40$. To test the sensitivity of the resulting image to the choice of $\alpha$ in \eref{E:TVP}, we performed reconstructions with $\alpha = 1.0, 0.8, 0.5, 0.2$, and $0.0$; the first value corresponding to the case of ordinary TV regularization. For comparison, we also reconstructed an image using the wPSART-TV approach of ~\citer{HG17}; this is equivalent to applying \aref{A:AlgwpSART-sup} with $x_0 = 0$ and $\phi = \phi_{TV}$.

For quantitative comparisons, we defined a $51 \times 51$ pixel region of interest (ROI) in the center of the object, surrounding the six small low-contrast features. Pixel values within the ROI were rescaled to the interval $[0,1]$ using the formula
\begin{equation}
x_{new,k} = \frac{x_{old,k}-x_{min}}{x_{max} - x_{min}},
\end{equation}
where $x_{min}$ and $x_{max}$ were set to $\pm 10\%$ of the background soft tissue value. The peak signal-to-noise ratio (PSNR) of the region was then calculated as
\begin{align}
PSNR &= 10 \log_{10} \left( \frac{1}{MSE} \right), \text{where} \\MSE &= \frac{1}{51^2}\sum_k \left(x_{true,k} - x_{new,k}\right)^2.
\end{align}
The rescaling was performed so that the PSNR values were better able to capture differences between the reconstructed values within the ROI.

\begin{figure}
\includegraphics[width=\linewidth]{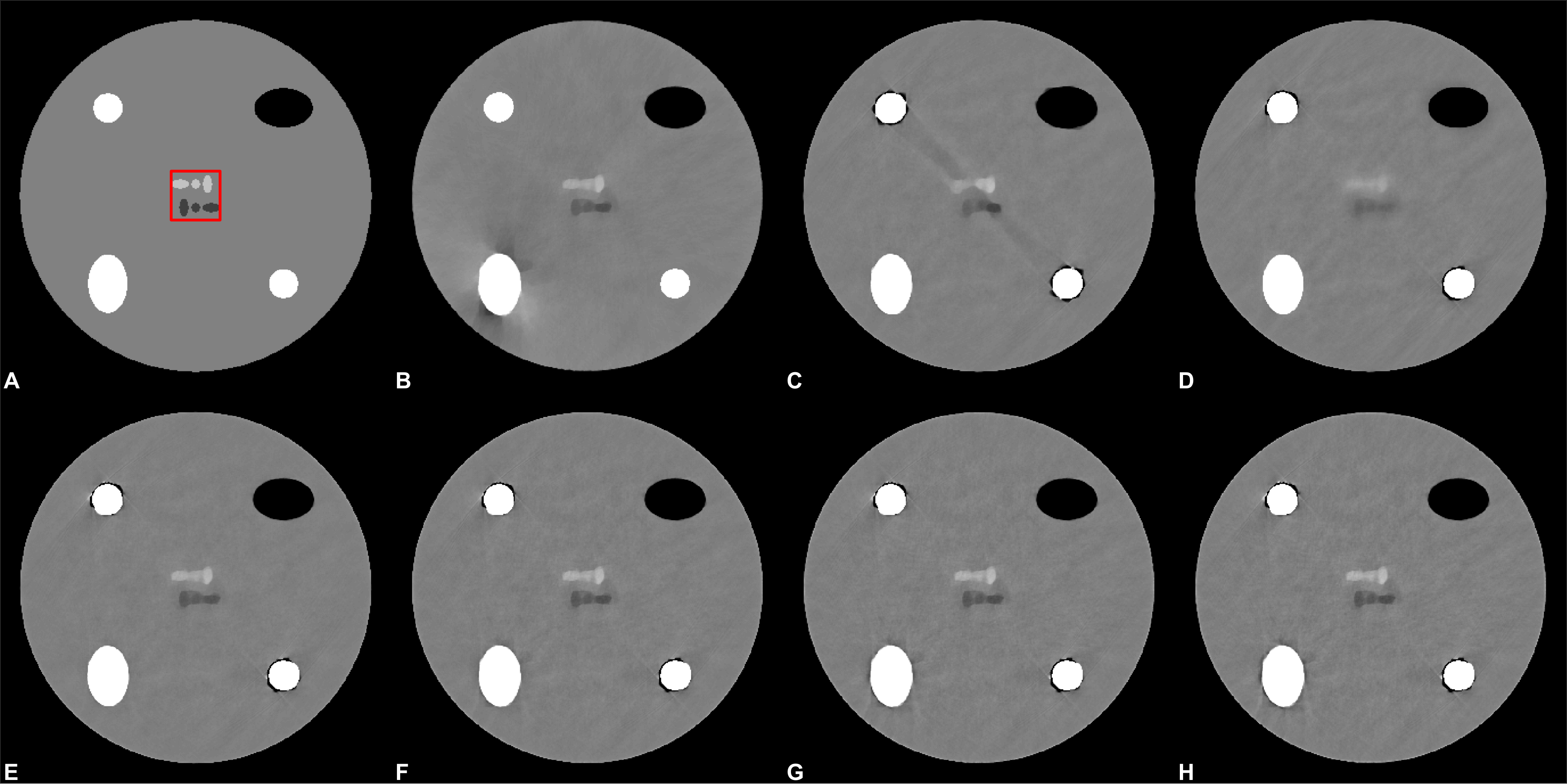}
\includegraphics[width=\linewidth]{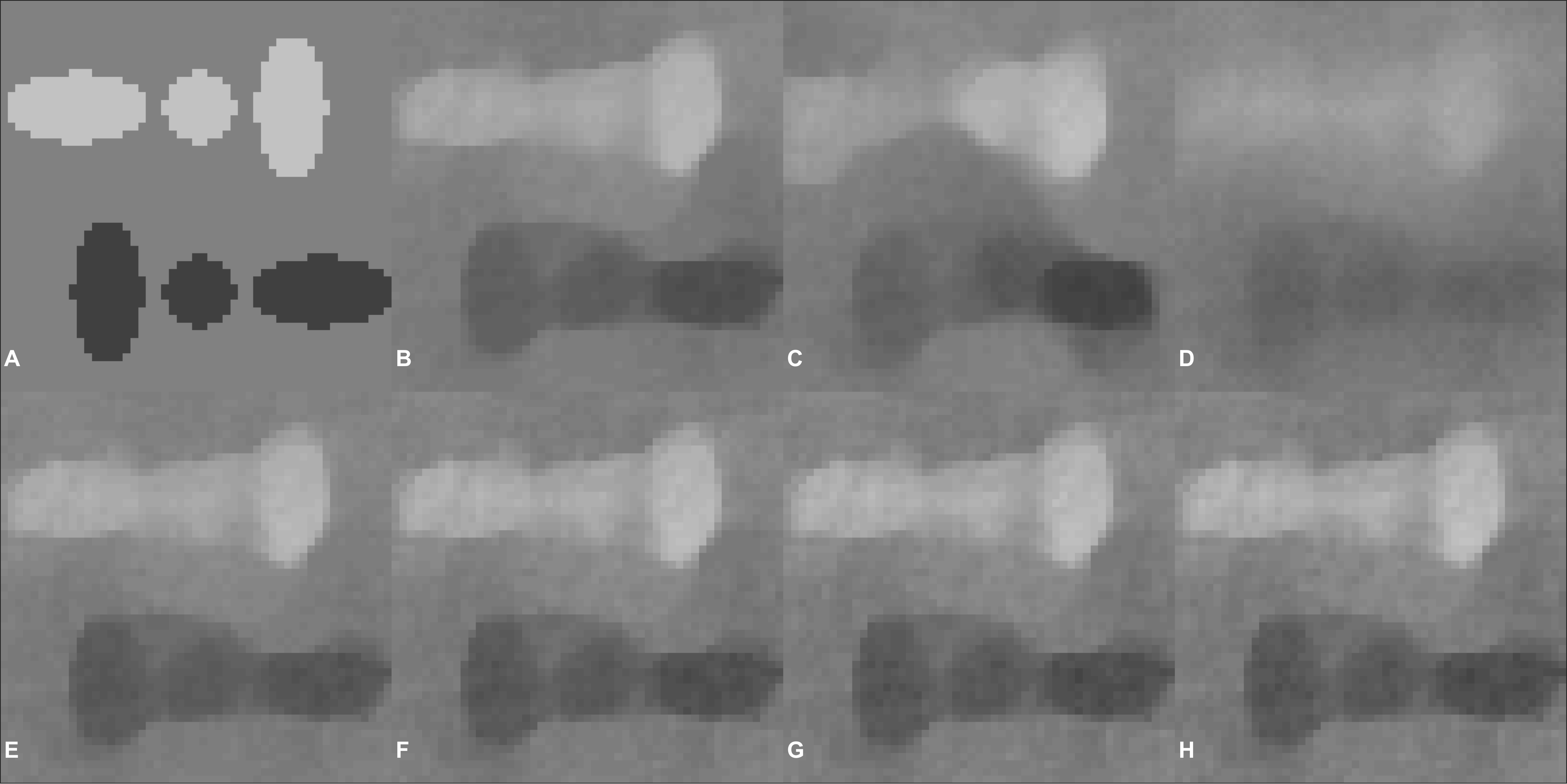}
\caption{Images reconstructed at noise level of $I_0 = 5.0 \times 10^5$. A: True image. B: Prior image $x_{prior}$ reconstructed from NMAR data. C: Image reconstructed using wPSART-TV. D -- H: Images reconstructed using wPSART-PICS with $\alpha = 1.0, 0.8, 0.5, 0.2$ and $0.0$, respectively. Top two rows: full image; bottom two rows: zoomed in on ROI highlighted in red box. Grayscale window is $\pm 10\%$ the value of the soft tissue background.}\label{F:simplephantom_comp}
\end{figure}

\fref{F:simplephantom_comp} shows the images reconstructed using all approaches at the noise level of $I_0 = 5.0 \times 10^5$. While the prior image reconstructed from the NMAR data (B) is free of metal streaking artifacts, there is a small artifact around the oblong bone region in the bottom right due to inaccuracy in the segmentation applied during NMAR. This artifact is absent from images C--H, which are reconstructed from the original data. Image C, reconstructed using the wPSART-TV algorithm, retains a mild streaking artifact between the two metal objects. This is likely due to the fact that the image is reconstructed without any prior knowledge from the NMAR correction, which removes the streak entirely. Image D begins from $x_0 = x_{prior}$, but uses only TV as the penalty function during superiorized reconstruction (since $\alpha = 1.0$), resulting in blurred edges around the low-contrast features. Images E--H, reconstructed using pSART-PICS with $\alpha < 1$, demonstrate better definition about the features.

The PSNR values given in \tref{T:simplephantom_comp} corroborate these observations. At all four noise levels, the best PSNR values are obtained using wPSART-PICS with an $\alpha$ value of 0.0 (column H). This indicates that (for this experiment, in any event), there is no benefit to including the TV of the image in the penalty function \eref{E:TVP}. This is perhaps not surprising, since the prior image itself is reconstructed using TV superiorization and is therefore quite smooth itself. That being said, the difference in PSNR values between columns E--H is only on the order of 1--2\%, indicating that the algorithm is not particularly sensitive to the choice of $\alpha$, as long as $\alpha < 1.0$. Comparatively, the difference in PSNR values between columns D and E ($\alpha = 1.0$ versus $\alpha = 0.8$) ranges from 2 -- 10\%.

\begin{table}
\caption{PSNR values of reconstructed images within the ROI shown in \fref{F:simplephantom_comp}. Column headings refer to images shown in \fref{F:simplephantom_comp}.}\label{T:simplephantom_comp}
\begin{tabular}{p{3.5em}|p{3em}p{3em}p{3em}p{3em}p{3em}p{3em}p{3em}}
Counts	&B &C &D &E &F &G &H \\
\hline
1.0e5	&21.36	&21.41	&21.24	&21.73	&21.92	&21.97	&21.97 \\
2.0e5	&21.96	&22.00	&21.30	&22.35	&22.49	&22.58	&22.62 \\
5.0e5	&23.28	&22.24	&21.37	&23.56	&23.82	&23.93	&23.97 \\
1.0e6	&23.95	&20.46	&21.67	&24.20	&24.52	&24.62	&24.66
\end{tabular}
\end{table}

\subsection{Anatomical phantom experiments}

\begin{figure}
\includegraphics[width=\linewidth]{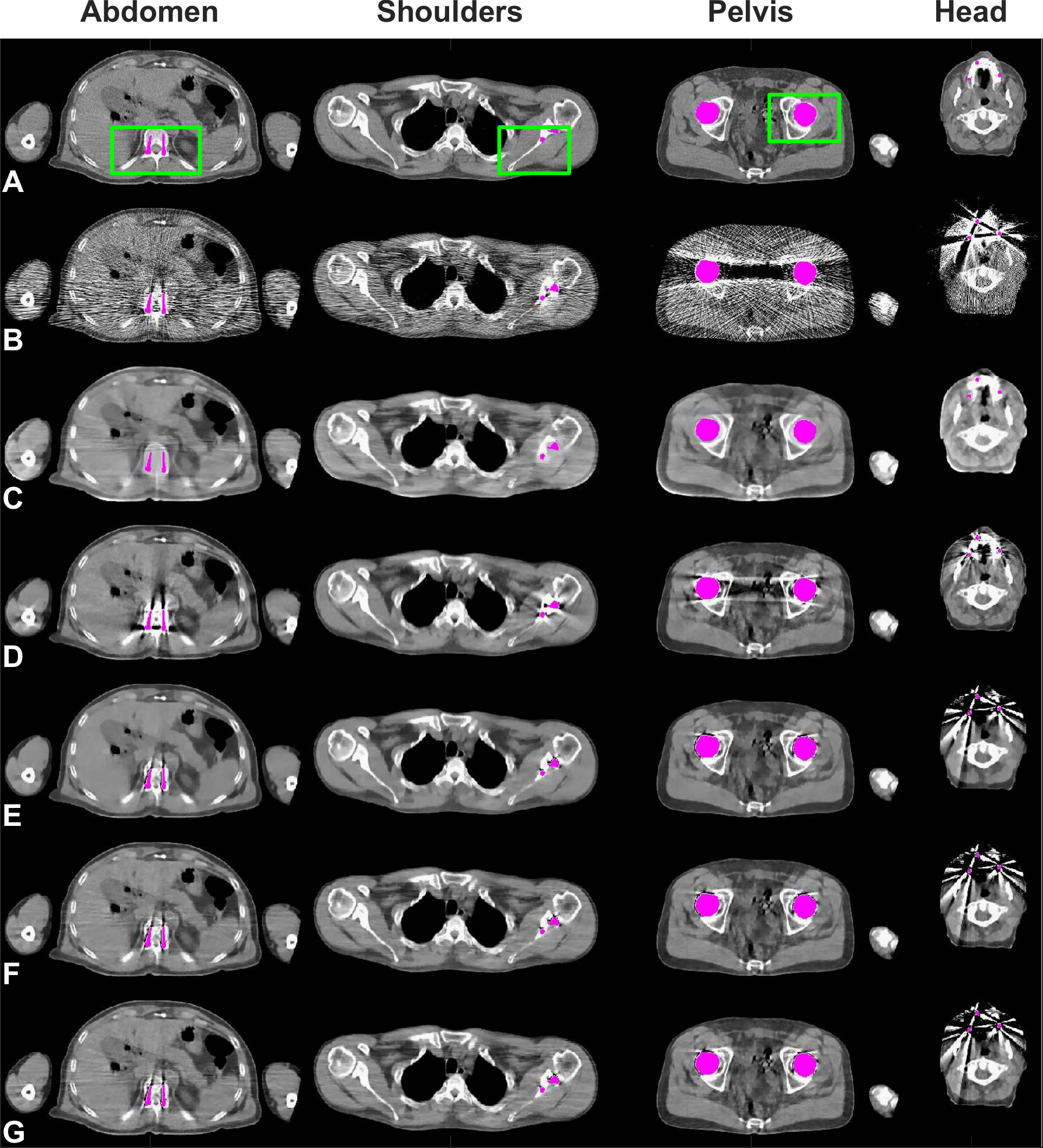}
\caption{Phantoms and reconstructions based on clinical CT image slices. Row A: True phantoms at reference energy of 70 keV with metal inserts indicated in pink. Row B: reconstructions generated using FBP and water corrected data. Row C: Prior images reconstructed from NMAR data. Row D: Images reconstructed using wPSART-TV. Rows E--G: Images reconstructed using wPSART-PICS with $\alpha = 1.0, 0.5$ and $0.0$, respectively. Grayscale window is $[0.15, 0.25]$ cm$^{-1}$. Images were reconstructed at 512$\times$512 pixels, but have been cropped to eliminate black space. Green boxes indicate zoomed-in regions shown in \fref{F:realdata_phantom_comp_zoom}.}\label{F:realdata_phantom_comp_full}
\end{figure}

To test the performance of the algorithm on more realistic data, slices of clinical CT images were downloaded from the Cancer Imaging Archive\footnote{https://www.cancerimagingarchive.net/}. Four $512 \times 512$ pixel slices were selected from the dataset, representing the abdominal, shoulder, pelvic and head regions. Metal objects representing pins, dental fillings, and a bilateral hip implant were subsequently manually inserted into the image slices; the pins and hip implants were modeled as titanium, while the fillings were modeled as gold. The objects were modeled after those appearing in \citer{ZY18}. Fan beam data corresponding to 900 views over 360$^\circ$ were simulated using the MIRT toolbox, using the same 130 kVp spectrum and reference energy of 70 keV as for the simple phantom experiment. Soft thresholding~\cite{ZY18} with base materials of fat, soft tissue, and bone were used to generate attenuation coefficients for nonmetal regions at every energy level when simulating the polyenergetic data. Noise was then added to the data proportional to count rate of $I_0 = 2.0 \times 10^5$. The four true images and their reconstructions using FBP are shown in the first two rows of \fref{F:realdata_phantom_comp_full}, illustrating the severity of the metal artifacts when no correction is performed.

Images were reconstructed using the same algorithms as in the previous section. In light of the results from the simple phantom experiment, we only used values of $\alpha = 1.0$, $0.5$ and $0.0$ for the pSART-PICS algorithm. All superiorized algorithms were run with $\gamma = 0.995$ and $N = 10$; while the choice of $\gamma = 0.9995$, $N = 40$ gave good results for the simple phantom used in the previous section, we found that it tended to oversmooth the images based on real CT data. By reducing both $\gamma$ and $N$, we smooth the images less because we perform fewer gradient descent iterations in \aref{A:AlgwpSART-sup}, while also reducing the step size more quickly.

\begin{figure}
\includegraphics[width=0.9\linewidth]{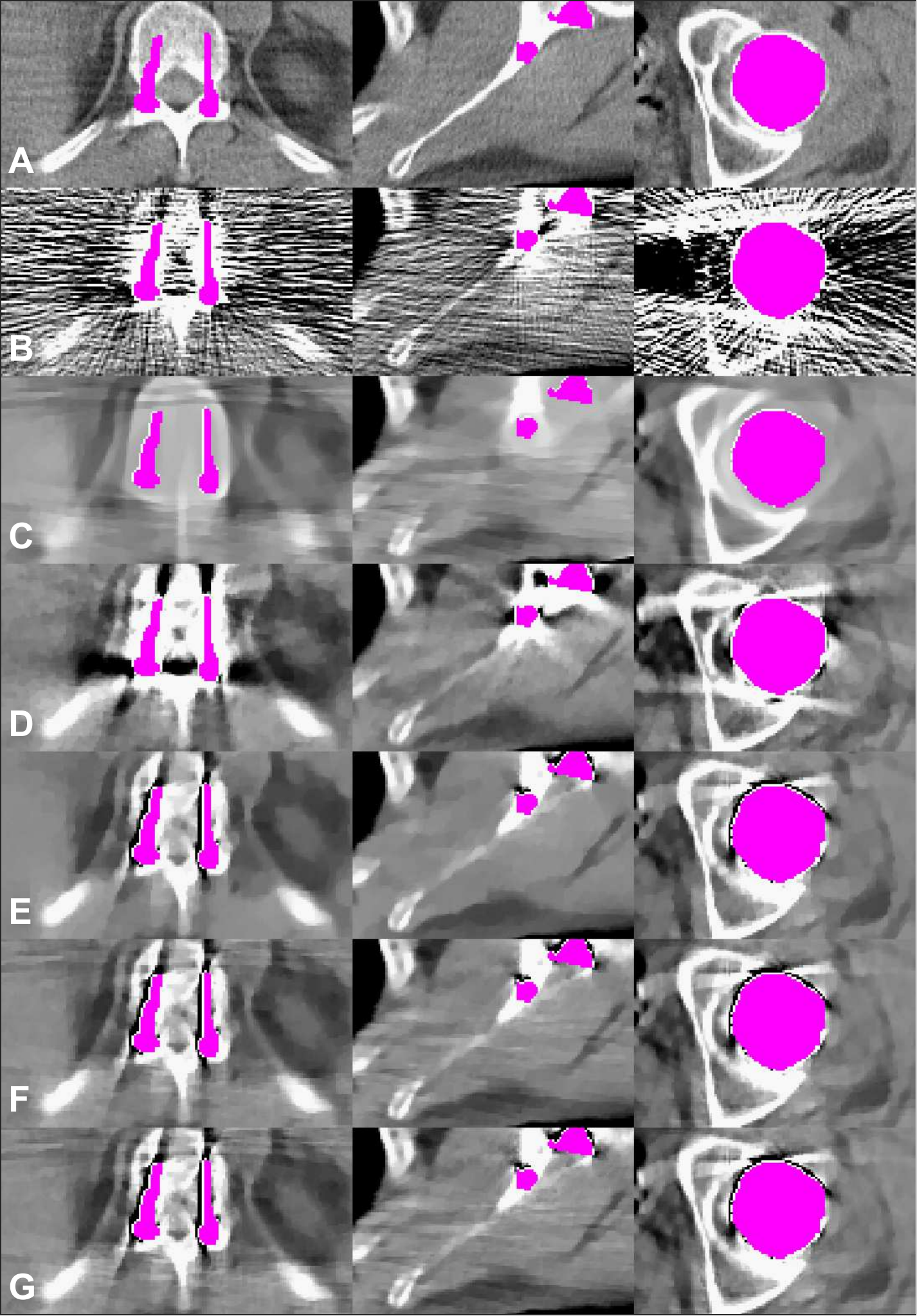}
\caption{Zoomed-in regions indicated in \fref{F:realdata_phantom_comp_full}.}\label{F:realdata_phantom_comp_zoom}
\end{figure}

Reconstructed images of the four phantoms are shown in \fref{F:realdata_phantom_comp_full} and \fref{F:realdata_phantom_comp_zoom}. Row C of the figures shows the prior images reconstructed from the NMAR data. It is apparent in  \fref{F:realdata_phantom_comp_zoom} that some information has been lost due to the interpolation performed by NMAR; for example, the area around the spine and backmost ribs in the abdomen image, the scapula on the right side of the shoulder image, and the edges of the hip bones in the pelvis image. While the wPSART-TV algorithm (Row D) does a better job of recovering these features, streaking artifacts caused by metal persist in every image, most noticeably between the two implants in the hip image. The three images reconstructed using wPSART-PICS (rows E--G) are able to eliminate streaks caused by metal while also recovering features which were occluded in the prior image. As in the simple phantom experiment, there is little apparent difference in the images reconstructed using $\alpha < 1.0$ (rows F and G), while the image using $\alpha = 1.0$ (i.e., TV as the penalty function) is noticeably smoother. In the abdomen and shoulder images, including the prior in the penalty function has actually introduced some mild streaking into the image along the horizontal direction, since these streaks are also present in the prior image.

It is apparent that the wPSART-PICS algorithm does not give an acceptable reconstructed image in the case of the head phantom (last column of \fref{F:realdata_phantom_comp_full}). The reason for this is that the metal simulated in the head phantom experiment (gold) has such a high $\mu$ value that the three objects modeling dental fillings block all X-rays passing through them at the simulated count rate of $2.0 \times 10^5$ counts per ray. This results in a measured intensity of $I_j = 0$, translating to infinite attenuation along the line. Due to the projection weighting employed by the algorithm~\eref{E:weight_matrix}, these measurements are ignored during reconstruction; however, the lack of information about pixel intensities along those lines produce artifacts in the final image. The artifacts are actually more severe in the images which begin with the prior as an initial estimate (Rows E--G); this is because there is a slight mismatch between the metal trace in the forward projection of the prior image and that of the measured sinogram, which creates severe streaking in the reconstructed image after just one iteration of the algorithm. 

\section{Conclusions}

In this paper we present an iterative reconstruction algorithm for CT imaging which uses the superiorization methodology to perform metal artifact reduction (MAR). The algorithm uses a prior image reconstructed using the normalized metal artifact reduction (NMAR) method of Meyer et al.~\cite{MRLSK10}, which eliminates most severe streaking artifacts caused by metal, but may introduce secondary artifacts in the vicinity of the metal regions. The NMAR-based prior is then incorporated into a penalty function akin to that used in prior image constrained compressive sensing (PICCS) algorithms~\cite{CTL08}, which is used within a superiorized iterative algorithm based on our previous work (wPSART-TV)~\cite{HG17}. Numerical experiments modeling several different anatomical scenarios were perfomed, and indicate that the proposed algorithm (wPSART-PICS) is able improve on both NMAR and wPSART-TV. In particular, it is able to recover details that are lost during the NMAR process (particularly with respect to the structure of bone around the metal regions) while also removing streaking artifacts that persist in images reconstructed using wPSART-TV.

\begin{figure}
\includegraphics[width= 0.6\linewidth]{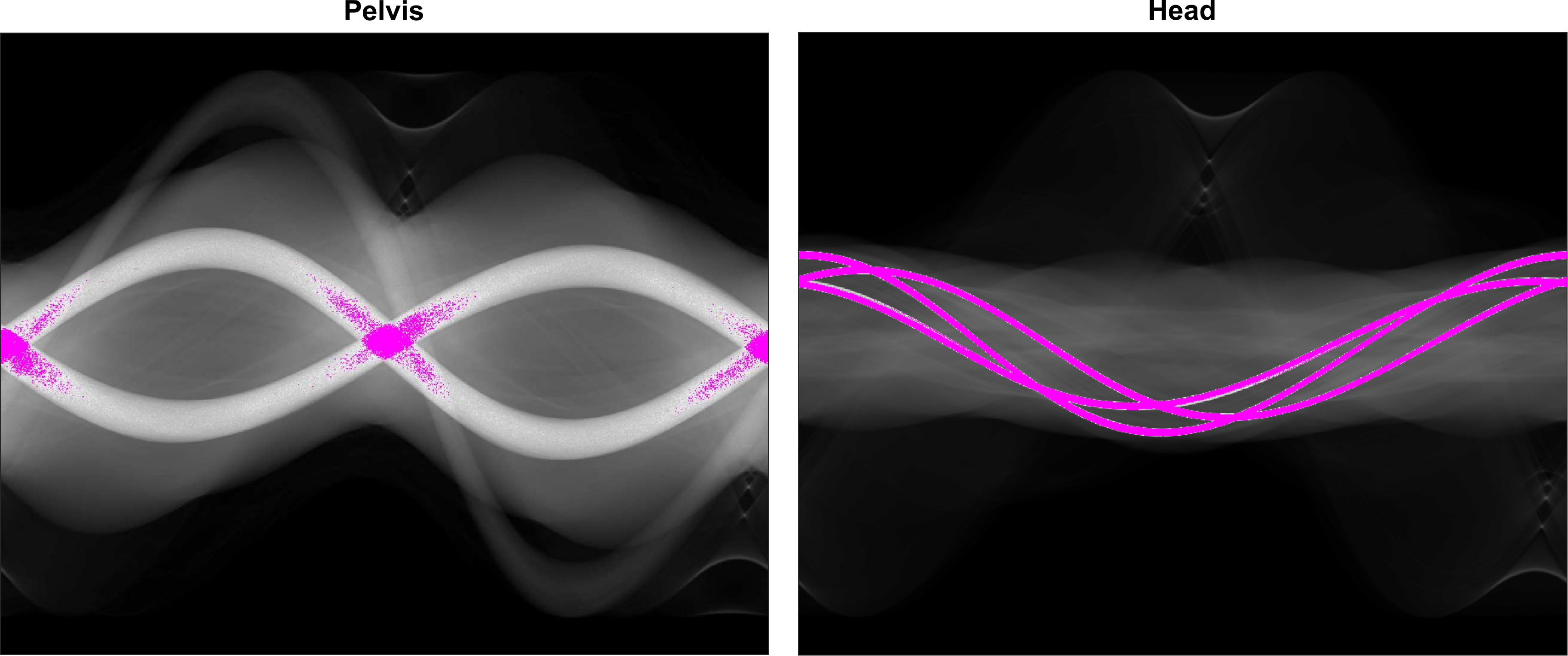}
\caption{Fan-beam sinograms corresponding to pelvis and head phantoms with initial count rate of $I_0 = 2.0 \times 10^5$. Measurements corresponding to infinite attenuation are shown in pink.}\label{F:pelvis_vs_head_sino}
\end{figure}

While the wPSART-PICS algorithm was effective in removing metal artifacts in most cases, the numerical experiments do highlight some limitations. In an experiment simulating gold dental fillings (fourth column of \fref{F:realdata_phantom_comp_full}), the algorithm failed to produce acceptable results, due to the total photon starvation induced by the dense metal objects. In a separate experiment (not shown), we simulated titanium fillings rather than gold; this eliminated the artifacts as the titanium fillings no longer blocked all X-rays. Additionally, we note that the simulated sinogram data for the pelvis phantom also included measurements along which no photons were detected; nonetheless, the wPSART-PICS algorithm was successfully able to reconstruct this image. As indicated in \fref{F:pelvis_vs_head_sino}, the key difference appears to be that not {\em all} measurements through the metal objects in the pelvis phantom are totally attenuated; only those passing through both objects. Thus, the algorithm is able to reconstruct the object accurately based on the remaining information. We conclude that while the wPSART-PICS algorithm is able to perform well in challenging scenarios including some photon starvation, the NMAR algorithm may be preferable in cases where extreme photon starvation occurs. Additional limitations include the use of spectral knowledge in the forward projection step~(\eref{E:fp_poly}), which may not always be available, and the need to tune parameters such as $N$, $\gamma$, $i_{max}$ and/or $\varepsilon$ in \aref{A:AlgwpSART-sup} to obtain good results.

%To address these different sources of error, we propose an iterative algorithm for CT image reconstruction which reduces metal artifacts. The algorithm does so by (1) accurately modeling polyenergetic X-ray data, (2) statistically weighting the X-ray data to reduce the effect of noisy measurements, and (3) incorporating total variation (TV) as a secondary objective.  Our numerical experiments indicate that all three of these features of the algorithm play an important role in reducing metal artifacts. The recently proposed superiorization methodology~\cite{HGDC12} provides a solid mathematical foundation for our approach.

\section*{Acknowledgments}
The authors would like to thank Aviv Gibali (ORT Braude College, Karmiel, Israel) for his initial suggestion to apply superiorization to metal artifact reduction, and contributions to earlier work.

\bibliographystyle{unsrt}
\bibliography{MAR}

%\begin{thebibliography}{10}
%
%\end{thebibliography}

\end{document}